\documentclass[]{aa}

\usepackage{graphicx}
\usepackage{amssymb}

\begin{document}

\title{Density Estimators in Particle Hydrodynamics}
\titlerunning{DTFE vs SPH}
\subtitle{DTFE versus regular SPH}

\author{Pelupessy F.I. \inst{1} \and
        Schaap W.E. \inst{2}      \and
        Van de Weygaert R.\inst{2}}
\authorrunning{ Pelupessy, Schaap, van de Weygaert}
\institute{ {Leiden Observatory, Leiden University, PO Box 9513, 2300 RA 
Leiden, The Netherlands \\ email: pelupes@strw.leidenuniv.nl}
 \and {Kapteyn Institute, University of Groningen, PO Box 800, 9700 AV 
Groningen, The Netherlands \\  email: wschaap@astro.rug.nl, 
weygaert@astro.rug.nl}}

\date{ }

\abstract{We present the results of a study confronting density
    maps reconstructed by the Delaunay Tessellation Field Estimator
    (DTFE) and by regular SPH kernel-based techniques. The density
    maps are constructed from the outcome of an SPH particle
    hydrodynamics simulation of a multiphase interstellar medium. The
    comparison between the two methods clearly demonstrates the
    superior performance of the DTFE with respect to conventional SPH
    methods, in particular at locations where SPH appears to fail.
    Filamentary and sheetlike structures form telling examples.  The
    DTFE is a fully self-adaptive technique for reconstructing
    continuous density fields from discrete particle distributions,
    and is based upon the corresponding Delaunay tessellation. Its
    principal asset is its complete independence of arbitrary smoothing
    functions and parameters specifying the properties of these.  As a
    result it manages to faithfully reproduce the anisotropies of the
    local particle distribution and through its adaptive and local 
    nature proves to be optimally suited for uncovering the full 
    structural richness in the density distribution. Through the 
    improvement in local density estimates, calculations invoking the
    DTFE will yield a much better representation of physical processes
    which depend on density. This will be crucial in the case of feedback
    processes, which play a major role in galaxy and star formation issues.
    The presented results form an encouraging step towards the application
    and insertion of the DTFE in astrophysical hydrocodes.  We
    describe an outline for the construction of a particle
    hydrodynamics code in which the DTFE replaces kernel-based
    methods.  Further discussion addresses the issue and possibilities
    for a moving grid based hydrocode invoking the DTFE, and Delaunay
    tessellations, in an attempt to combine the virtues of the
    Eulerian and Lagrangian approaches.  }

\maketitle

\section{Introduction}

Smoothed Particle Hydrodynamics (SPH) has established itself as the
workhorse for a variety of astrophysical fluid dynamical
computations (Lucy~\cite{Lc77}, Ginghold \& Monaghan~\cite{GinMon77}).
In a wide range of astrophysical environments this Lagrangian scheme
offers substantial and often crucial advantages over Eulerian, usually
grid-based, schemes. Astrophysical applications such as
cosmic structure formation and galaxy formation, the dynamics of
accretion disks and the formation of stars and planetary systems are
testament to its versatility and succesful
performance (for an enumeration of applications, and corresponding
references, see e.g. the reviews by Monaghan~\cite{Mon92} and
Bertschinger~\cite{Ber98}).

A crucial aspect of the SPH procedure concerns the proper estimation
of the local density, i.e. the density at the location of the
particles which are supposed to represent a fair -- discrete --
sampling of the underlying continuous density field. The basic feature
of the SPH procedure for density estimation is based upon a
convolution of the discrete particle distribution with a particular
user-specified kernel function $W$. For a sample of $N$ particles,
with masses $m_j$ and locations ${\bf r}_j$, the density $\rho$ at the
location ${\bf r}_i$ of particle $i$ is given by
\begin{equation}
\label{sphdensity} \rho({\bf r}_i)\,=\,\sum_{j=1}^{N} \,m_j\,
W({\bf r}_i-{\bf r}_j,h_i)\,,
\end{equation}
in which the kernel resolution is determined through the smoothing
scale $h_i$. Notice that generically the scale $h_i$ may be different
for each individual particle, and thus may be set to adapt to the
local particle density. Usually the functional dependence of the
kernel $W$ is chosen to be spherically symmetric, so that it is a
function of $|{\bf r}_i-{\bf r}_j|$ only. 

The evolution of the physical system under consideration is fully
determined by the movement of the discrete particles. Given a properly
defined density estimation procedure, the equations of motion for the
set of particles are specified through a suitable Lagrangian, if
necessary including additional viscous forces (see e.g.~Rasio \cite{Ras99}).

In practical implementations, however, the SPH procedure involves a
considerable number of artefacts. These stem from the fact that SPH
particles represent functional averages over a certain Lagrangian
volume. This averaging procedure is further aggravated by the fact
that it is based upon a rather arbitrary user-specified choice of both
the adopted resolution scale(s) $h_i$ and the functional form of the
kernel $W$. Such a description of a physical system in terms of
user-defined fuzzy clouds of matter is known to lead to considerable
complications in realistic astrophysical circumstances. Often, these
environments involve fluid flows exhibiting complex spatial patterns
and geometries. In particular in configurations characterized by
strong gradients in physical characteristics -- of which the density,
pressure and temperature discontinuities in and around shock waves
represent the most frequently encountered example -- SPH has been
hindered by its relative inefficiency in resolving these gradients.

Given the necessity for the user to specify the characteristics and
parameter values of the density estimation procedure, the accuracy and
adaptibility of the resulting SPH implementation hinges on the ability
to resolve steep density contrasts and the capacity to adapt itself to
the geometry and morphology of the local matter distribution. A
considerable improvement with respect to the early SPH
implementations, which were based on a uniform smoothing length $h$,
involves the use of adaptive smoothing lengths $h_i$ (Hernquist \&
Katz \cite{HK89}), which provides the SPH calculations with a larger
dynamic range and higher spatial resolution. The mass distribution in
many (astro)physical systems and circumstances is often characterized
by the presence of salient anisotropic patterns, usually identified as
filamentary or planar features. To deal with such configurations,
additional modifications in a few sophisticated implementations
attempted to replace the conventional -- and often unrealistic and
restrictive -- spherically symmetric kernels by ones whose
configuration is more akin to the shape of the local mass
distribution. The corresponding results do indeed represent a strong
argument for the importance of using geometrically adaptive density
estimates.  A noteworthy example is the introduction of ellipsoidal
kernels by Shapiro et al.~(\cite{SM96}). Their shapes are stretched in
accordance with the local flow. Yet, while evidently being
conceptually superior, their practical implementation does constitute
a major obstacle and has prevented widescale use. This may be ascribed
largely to the rapidly increasing number of degrees of freedom needed
to specify and maintain the kernel properties during a simulation.

Even despite their obvious benefits and improvements, these methods
are all dependent upon the artificial parametrization of the local
spatial density distribution in terms of the smoothing kernels.
Moreover, the specification of the information on the density
distribution in terms of extra non physical variables, necessary for
the definition and evolution of the properties of the smoothing
kernels, is often cumbersome to implement and may introduce subtle
errors (Hernquist \cite{H93}, see however Nelson \& Papaloizou
\cite{NP94} and Springel \& Hernquist \cite{SH02}).  In many
astrophysical applications this may lead to systematic artefacts in
the outcome for the related physical phenomena. Within a cosmological
context, for example, the X-ray visibility of clusters of galaxies is
sensitively dependent upon the value of the local density, setting the
intensity of the emitted X-ray emission by the hot intergalactic gas.
This will be even more critical in the presence of feedback processes,
which for sure will be playing a role when addressing the amount of
predicted star formation in simulation studies of galaxy formation.

Here, we seek to circumvent the complications induced by the kernel
parametrization and introduce and propose an alternative to the use of
kernels for the quantification of the density within the SPH
formalism. This new method, based upon the Delaunay Tessellation Field
Estimator (DTFE, Schaap \& van de Weygaert \cite{SW00}), has been
devised to mould and fully adapt \emph{itself} to the configuration of
the particle distribution. Unlike conventional SPH methods, it is able
to deal self-consistently and naturally with anisotropies in the
matter distribution, even when it concerns caustic-like transitions.
In addition, it manages to succesfully treat density fields marked by
structural features over a vast (dynamic) range of scales.

The DTFE produces density estimates on the basis of the particle
distribution, which is supposed to form a discrete spatial sampling of
the underlying continuous density field. As a linear multidimensional
field interpolation algorithm it may be regarded as a first-order
version of the natural neighbour algorithm for spatial interpolation
(Sibson \cite{Sib81}, also see e.g.~Okabe et al.~\cite{Oketal00}). In
general, applications of the DTFE to spatial point distributions have
demonstrated its success in dealing with the complications of
anisotropic geometry and dynamic range (Schaap \& van de Weygaert
\cite{SW00}). The key ingredient of the DTFE procedure is that of the
Delaunay triangulation, serving as the complete covering of a sample
volume by mutually exclusive multidimensional linear interpolation
intervals.

Delaunay tessellations (Delone 1934; see e.g. Okabe et al. 2000 for 
extensive review) form the natural framework in which to discuss
the properties of discrete point sets, and thus also of discrete
samplings of continuous fields. Their versatility and significance 
have been underlined by their widespread applications in such areas 
as computer graphics, geographical mapping and medical imaging. Also, 
they have already found widespread application in a variety of 
`conventional'grid-based fluid dynamical computation schemes. This may 
concern their use as a non-regular application-oriented grid covering of 
physicalsystems, which represents a prominent procedure in technological
applications. More innovating has been their use in Lagrangian
`moving-grid' implementations (see Mavripilis \cite{Mv97} for a
review, and Whitehurst \cite{Wh95} for a promising astrophysical
application).

It seems therefore a good idea to explore the possibilities of
applying the DTFE in the context of a numerical hydrodynamics code.
Here, as a first step, we wish to obtain an idea of the performance of
a hydro code involving the use of DTFE estimates with respect to an
equivalent code involving regular SPH density estimates. The quality
of the new DTFE method with respect to the conventional SPH estimates,
and their advantages and disadvantages under various circumstances,
are evaluated by a comparison between the density field which would be
yielded by a DTFE processing of the resulting SPH particle
distribution and that of the regular SPH procedure itself. In this
study, we operate along these lines by a comparison of the resulting
matter distributions in the situation of a representative stochastic
multiphase density field.  This allows us to make a comparison between
both density estimates in a regime for which an improved method for
density estimates would be of great value. We should point out a major
drawback of our approach, in that we do not really treat the DTFE
density estimate in a self-consistent fashion. Instead of being part
of the dynamical equations themselves we only use it as an analysis
tool of the produced particle distribution. Nevertheless, it will
still show the value of the DTFE in particle gasdynamics and give an
indication of what kind of differences may be expected when
incorporating in a fully self-consistent manner the DTFE estimate in
an hydrocode.

On the basis of our study, we will elaborate on the potential benefits
of a hydrodynamics scheme based on the DTFE. Specifically, we outline
how we would set out to develop a complete particle hydrodynamics code
whose artificial kernel based nature is replaced by the more natural
and self-adaptive approach of the DTFE. Such a DTFE based particle
hydrodynamics code would form a promising step towards the development
of a fully tessellation based quasi-Eulerian moving-grid
hydrodynamical code. Such would yield a major and significant step
towards defining a much needed alternative and complement to currently
available simulation tools.

\section{DTFE and SPH density estimates}
The methods we use for SPH and DTFE density estimates have been
extensively described elsewhere (Hernquist \& Katz \cite{HK89}, Schaap
\& van de Weygaert \cite{SW00}). Here, we will only summarize their
main, and relevant, aspects.

\begin{figure*}[t]
\centering
\includegraphics[width=17.cm]{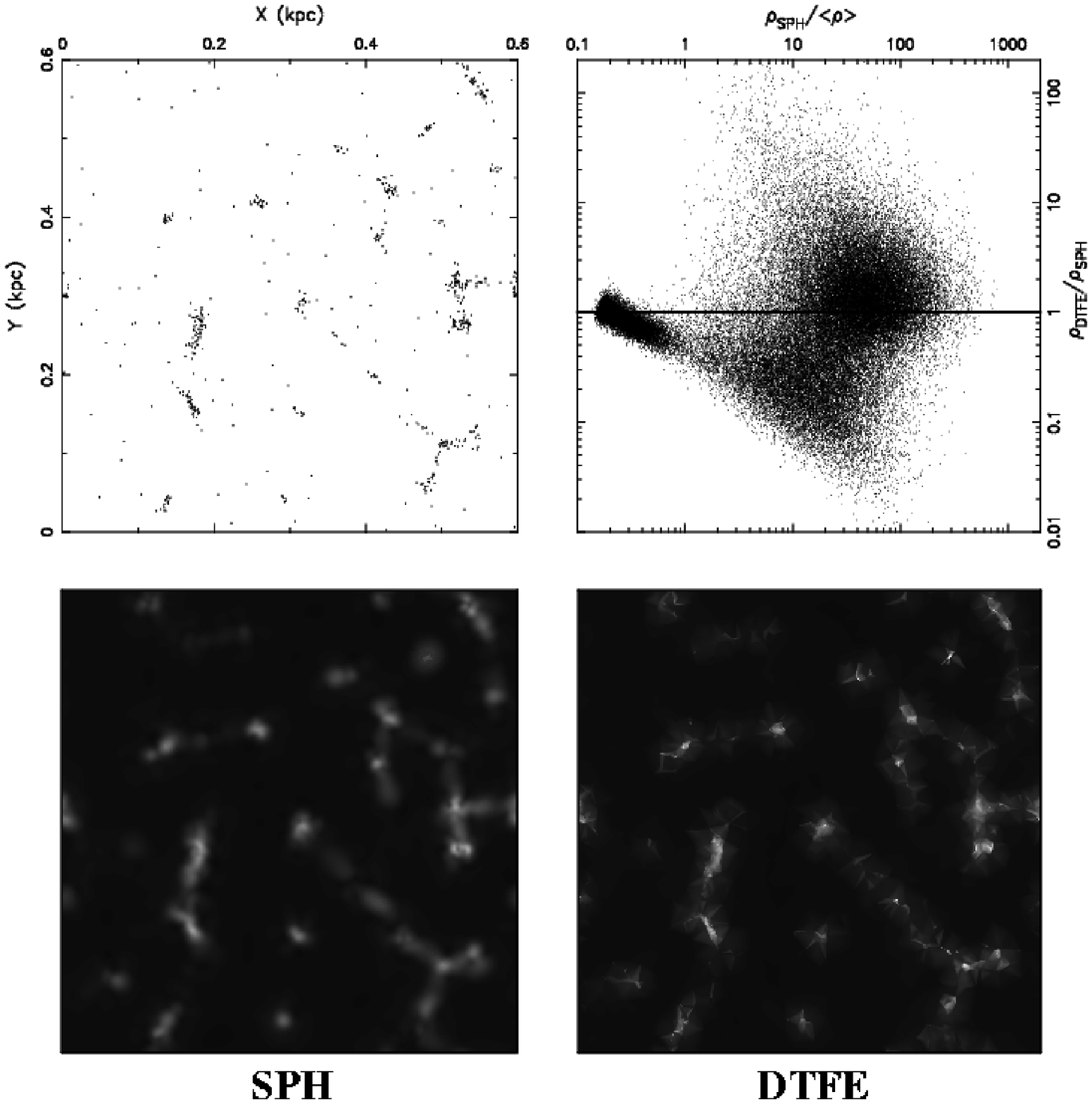}
\caption{Comparison of the DTFE performance versus that of the regular 
SPH method in a characteristic configuration, that of a hydrodynamic 
simulation of the multiphase interstellar medium. Top left panel: the particle 
distribution in a $0.6\times 0.6$ kpc simulation region, within a slice with a 
width of 0.005 kpc. Bottom left frame: 2-D slice through the resulting (3-D) SPH 
density field reconstruction. Bottom right frame: the corresponding (3-D) density field 
reconstruction produced by the DTFE procedure. Top righthand frame: summary, 
in terms of a quantitative point-by-point comparison between the DTFE and SPH 
density estimates, $\rho_{\scriptstyle{\rm DTFE}}$ and 
$\rho_{\scriptstyle{\rm SPH}}$. Abscissa: the value of the SPH density 
estimate (normalized by the average density $\langle \rho \rangle$). Ordinate: the 
ratio of DTFE estimate to the SPH density estimate, 
$\rho_{\scriptstyle{\rm DTFE}}/\rho_{\scriptstyle{\rm SPH}}$. These 
quantities are plotted for each particle location in the full simulation 
box.}
 \label{figure1}
\end{figure*}

\subsection{SPH density estimate}
Amongst the various density recepies employed within available SPH
codes, we use the Hernquist \& Katz (\cite{HK89}) symmetrized form of
Eq. \ref{sphdensity}, using adaptive smoothing lengths:
\begin{equation}
\label{hkdensity} {\hat \rho}_i\,=\,\frac{1}{2}\,\sum_{j}\,m_j\,\left\{W(|r_i-r_j|,h_i)+
W(|r_i-r_j|,h_j)\right\}\,,
\end{equation}
\noindent The smoothing lengths $h_i$ are chosen such that the sum 
involves around 40 nearest neighbours. For the kernel $W$ we take the
conventional spline kernel described by Monaghan (\cite{Mon92}). Other
variants of the SPH estimate produce comparable results.

\subsection{DTFE density estimate}
The DTFE density estimating procedure consists of three basic steps. 

Starting from the sample of particle locations, the first step
involves the computation of the corresponding Delaunay tessellation.
Each Delaunay cell $T_m$ is the uniquely defined tetrahedron whose
four vertices (in 3-D) are the set of 4 sample particles whose
circumscribing sphere does not contain any of the other particles in
the set. The Delaunay tessellation is the full covering of space by
the complete set of these mutually disjunct tetrahedra. Delaunay
tessellations are well known concepts in stochastic and computational
geometry (Delaunay \cite{Del34}, for further references see e.g. Okabe
et al.~\cite{Oketal00}, M{\o}ller~\cite{Mol94} and van de Weygaert
\cite{Wey91}). 

The second step involves estimating the density at the location of
each of the particles in the sample. From the definition of the
Delaunay tessellation, it may be evident that there is a close
relationship between the volume of a Delaunay tetrahedron and the
local density of the generating point process (telling examples of
this may be seen in e.g. Schaap \& van de Weygaert 2002a). Evidently,
the ``empty'' cirumscribing spheres corresponding to the Delaunay
tetrahedra, and the volumes of the resulting Delaunay tetrahedra, will be 
smaller as the number density of sample points increases, and vice versa. 
Following this observation, a proper density estimate ${\hat \rho}$ at the location
${\bf x}_i$ of a sampling point $i$ is obtained by determining the
properly calibrated inverse of the volume ${\cal W}_{\scriptstyle{\rm
    Vor},i}$ of the corresponding {\it contiguous Voronoi cell}. The
contiguous Voronoi cell ${\cal W}_{\scriptstyle{\rm Vor},i}$ is the
union of all Delaunay tetrahedra $T_{m,i}$ of which the particle $i$
forms one of the four vertices, i.e. 
${\cal W}_{\scriptstyle{\rm Vor},i}=\bigcup_m T_{m,i}$. In general, when 
a particle $i$ is surrounded by $N_{T}$ Delaunay tetrahedra, each with a 
volume ${\cal V}({T_{m,i}})$, the volume of the resulting contiguous Voronoi 
cell is
\begin{equation}
\label{eq: contigvor}
{\cal W}_{\scriptstyle{\rm Vor},i}\,=\,\sum_{m=1}^{N_T} {\cal V}({T_{m,i}})\,.
\end{equation}
\noindent Note that $N_T$ is not a constant, but in general may acquire 
a different value for each point in the sample. For a Poisson
distribution of particles this is a non-integer number in the order of
$\langle N_{T}\rangle\approx 27$ (van de Weygaert~\cite{Wey94}).
Generalizing to an arbitrary $D$-dimensional space, and assuming that
each particle $i$ has been assigned a mass $m_i$, the estimated
density ${\hat \rho}_i$ at the location of particle $i$ is given by
(see Schaap \& van de Weygaert~\cite{SW00})
\begin{equation}\label{eq: rhofirst}
\label{DTFEdensity}{\hat \rho}({\bf r}_i)\,=\,(D+1)\,\frac{m_i}{{\cal W}_{\scriptstyle{\rm Vor},i}}\,,
\end{equation}
\noindent In this, we explicitly express ${\cal W}_{\scriptstyle{\rm Vor},i}$ 
for the general $D$-dimensional case. The factor $(D+1)$ is a
normalization factor, accounting for the $(D+1)$ different contiguous
Voronoi hypercells to which each Delaunay hyper``tetrahedron'' is
assigned, one for each vertex of a Delaunay hyper``tetrahedron''.

The third step is the interpolation of the estimated densities ${\hat
  \rho}_i$ over the full sample volume. In this, the DTFE bases itself
upon the fact that each Delaunay tetrahedron may be considered the
natural multidimensional equivalent of a linear interpolation interval
(see e.g Bernardeau \& van de Weygaert \cite{BerWey96}). Given the
$(D+1)$ vertices of a Delaunay tetrahedron with corresponding density
estimates ${\hat \rho}_j$, the value ${{\hat \rho}({\bf r})}$ at any
location ${\bf r}$ within the tetrahedron can be straightforwardly
determined by simple linear interpolation,
\begin{equation}
{\hat \rho}({\bf r})\,=\,{\hat \rho}({\bf r}_{i0})\,+\,
({\hat {\nabla \rho}})_{\scriptstyle {\rm Del},m}\,\cdot\,({\bf r}-{\bf r}_{i0})\,,  
\end{equation}
\noindent in which ${\bf r}_{i0}$ is the location of one of the Delaunay 
vertices $i$. This is a trivial evaluation once the value of the
(linear) density gradient $({\hat {\nabla \rho}})_{\scriptstyle {\rm
    Del},m}$ has been estimated.  For each Delaunay tetrahedron $T_m$
this is accomplished by solving the the system of $D$ linear equations
corresponding to each of the remaining $D$ Delaunay vertices
constituting the Delaunay tetrahedron $T_m$.  The ``minimum
triangulation'' property of Delaunay tessellations underlying this
linear interpolation, minimum in the sense of representing a
volume-covering network of optimally compact multidimensional
``triangles'', has been a well-known property utilized in a variety of
imaging and surface rendering applications such as geographical
mapping and various computer imaging algorithms.

\subsection{Comparison}
Comparing the two methods, we see that in the case of SPH the particle
`size' and `shape' (i.e.~its domain of influence) is determined by
some arbitrary kernel $W(r,h_i)$ and a fortuitous choice of smoothing
length $h_i$ (assuming, along with the major share of SPH procedures,
a radially symmetric kernel). In the case of the DTFE method the
particles' influence region is fully determined by the sizes and shapes 
of the Delaunay cells $T_{m,i}$, themselves solely dependent on the particle 
distribution. In other words, in regular SPH the density is determined 
through the kernel function $W({\bf x})$, while in DTFE it is solely the particle 
distribution itself setting the estimated values of the density. Contrary to the
generic situation for the kernel dependent methods, there are no extra
variables left to be determined. One major additional advantage is
that it is therefore not necessary to worry about the evolution of the
kernel parameters.

Both methods do display some characteristic artefacts in their density
reconstructions (see Fig.~1). To a large extent these may be traced back to the
implicit assumptions involved in the interpolation procedures, a
necessary consequence of the finite amount of information contained in
a discrete representation of a continuous field.  SPH density fields
implicitly contain the imprint of the specified and applied kernel
which, as has been discussed before, may seriously impart its
resolving power and capacity to trace the true geometry of structures.
The DTFE technique, on the other hand, does produce triangular
artefacts. At instances conspicuously visible in the DTFE
reconstructed density fields, they are the result of the linear
interpolation scheme employed for the density estimation at the
locations not coinciding with the particle positions. In principle,
this may be substantially improved by the use of higher order
interpolation schemes. Such higher-order schemes have indeed been
developed, and the ones based upon the natural neighbour interpolation
prescription of Sibson (\cite{Sib81}) have already been succesfully
applied to two-dimensional problems in the field of geophysics
(Sambridge et al.~\cite{Sametal95}, Braun \&
Sambridge~\cite{BraSam95}) and solid state physics
(Sukumar~\cite{Suk98}).

\begin{figure*}
\centering
\vskip 0.25truecm
\mbox{\hskip 0.5truecm\includegraphics[width=16.cm]{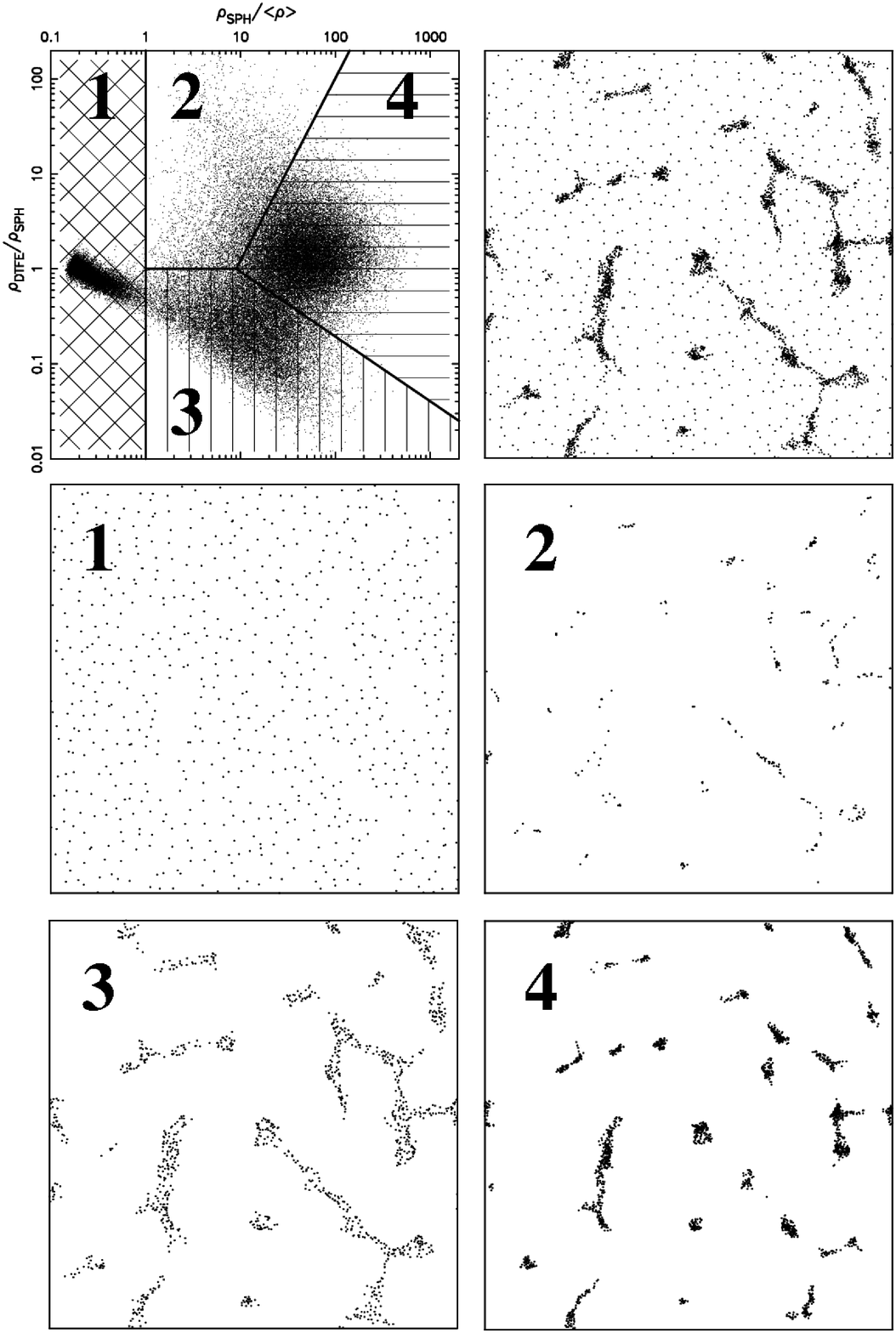}}
\label{figure2}
\end{figure*}
\begin{figure*}
\caption{Systematic analysis of the differences between the DTFE and 
SPH density estimates, $\rho_{\scriptstyle{\rm DTFE}}$ and 
$\rho_{\scriptstyle{\rm SPH}}$. 
Basis of the analysis is a point-by-point comparison of these two density 
estimates. Top lefthand frame: diagram of the value of the ratio 
$\rho_{\scriptstyle{\rm DTFE}}/\rho_{\scriptstyle{\rm SPH}}$ (ordinate) 
versus $\rho_{\scriptstyle{\rm SPH}}/\langle \rho \rangle$ 
(abscissa) for each of the points in the 
simulation volume. Indicated in this scatter diagram are four sectors, each of 
which corresponds to particles residing in a physically 
different regime/phase. On the basis of this identification, the full set of 
particles is dissected into the corresponding four composing particle samples.  
Top righthand frame: the spatial distribution of the full set of particles 
in a 0.04 kpc wide slice. The subsequent 4 frames (from central left to bottom 
right) show, for each indicated sector in the scatter diagram, the spatial distribution 
of the corresponding particles (within the same 0.04 kpc slice).}
\end{figure*}

\section{Case study: two-phase interstellar medium}
For the sake of testing and comparing the SPH and DTFE methods, we
assess a snapshot from a simulation of the neutral ISM. The model of
the ISM is chosen as an illustration rather than as a realistic model.

The ``simulation'' sample of the ISM consists of HI gas confined in a
periodic simulation box with a size $L=0.6~$kpc$^3$. The initially
uniform density of the gas is $n_H=0.3$ cm$^{-3}$, while its
temperature is taken to be $T=10000~$K. No fluctuation spectrum 
is imposed to set the initial featureless spatial gas distribution. 
To set the corresponding initial spatial distribution of the 
$N=64000$ simulation particles, we start from relaxed initial 
conditions according to a ``glass'' distribution (e.g. White~\cite{white94}). 

The evolution of the gas is solely a consequence of fluid dynamical
and thermodynamical processes. No self gravity is included. As for the
thermodynamical state of the gas, cooling is implemented using a fit
to the Dalgarno-McCray (\cite{DalCra72}) cooling curve. The heating of
the gas is accomplished through photo-electric grain heating,
attributed to a constant FUV background ($1.7$~G$_0$, with G$_0$ the
Habing field) radiation field. The parameters are chosen such that
after about 15 Myrs a two-phase medium forms which consists of warm
($10000$~K) and cold ($>100$~K) HI gas.

The stage at which a two-phase medium emerges forms a suitable point
to investigate the performance of the SPH and DTFE methods. At this
stage we took a snapshot from the simulation, and subjected it to
further analysis. For a variety of reasons, the spatial gas
distribution of the snapshot is expected to represent a challenging
configuration. The multiphase character of the resulting particle
configuration is likely to present a problem for regular SPH. Density
contrasts of about four orders of magnitude separate dense clumps from
the surrounding diffuse medium through which they are dispersed. Note
that a failure to recover the correct density may have serious
repercussions for the computed effects of cooling.  In addition, we
notice the presence of physical structures with conspicuous,
aspherical geometries (see Fig.~1 \& 2), such as anisotropic sheets
and filaments as well as dense and compact clumps, which certainly do
form a challenging aspect for the different methods.

\subsection{Results}
Fig.~\ref{figure1} offers a visual impression of the differences in
performance between the SPH and DTFE density reconstructions.  The
greyscale density maps in Fig.~1 (lower left: SPH, lower right: DTFE)
represent 2-D cuts through the corresponding 3-D density field
reconstructions (note that contrary to the finite width of the
corresponding particle slice, upper left frame, these constitute
planes with zero thickness).

Immediately visible is the more crispy appearance of the DTFE density
field, displaying substantially more contrast in conjunction with more
pronounced structural features. Look e.g. at the compact clump in the
lower righthand corner ($X\approx0.5,Y\approx0.12$), forming a
prominent and tight spot in the DTFE density field. The clump at
($X\approx0.48,Y\approx0.52$) represents another telling example,
visible as a striking peak in the DTFE rendering while hardly
noticeable in the SPH reconstruction. Structures in the SPH field have
a more extended appearance than their counterparts in the DTFE field,
whose matter content has been smeared out more evenly, over a larger
volume, yielding features with a significantly lower contrast. In this
assessment it becomes clear that the DTFE reconstruction adheres
considerably closer to the original particle distribution (top
lefthand frame). Apparently the DTFE succeeds better in rendering the
shapes, the coherence and the internal composition in the displayed
particle distribution. At various locations, the DTFE even manages to
capture structural details which seem to be absent in the SPH density
field.

To quantify the visual impressions of Fig.~1, and to analyze the
nature of the differences between the two methods, we plot the ratio
$\rho_{\scriptstyle{\rm DTFE}}/\rho_{\scriptstyle{\rm SPH}}$ as a
function of the SPH density estimate $\rho_{\scriptstyle{\rm
    SPH}}/\langle\rho\rangle$ (in units of the average density
$\langle \rho \rangle$).  Doing so for all particles in the sample
(Fig.~1, top righthand, Fig.~2, top lefthand) immediately reveals
interesting behaviour. The scatter diagram does show that the
discrepancies between the two methods may be substantial, with density
estimates at various instances differing by a factor of 5 or more.

Most interesting is the finding that we may distinguish clearly
identifiable and distinct regimes in the scatter diagram of
$\rho_{\scriptstyle{\rm DTFE}}/\rho_{\scriptstyle{\rm SPH}}$ versus
$\rho_{\scriptstyle{\rm SPH}}/\langle\rho\rangle$. Four different
sectors may be identified in the scatter diagram. Allowing for some
arbitrariness in their definition, and indicating these regions by
digits 1 to 4, we may organize the particles according to
density-related criteria, roughly specified as (we refer to Fig.~2,
top left frame, for the precise definitions of the domains):
\begin{enumerate}
\item low density regions: \\
\hspace*{0.0cm} $\rho_{\scriptstyle{\rm SPH}}/\langle\rho\rangle\,<\,1$
\item medium density regions, DTFE smaller than SPH: \\
\hspace*{0.0cm} $\rho_{\scriptstyle{\rm DTFE}}\,<\,\rho_{\scriptstyle{\rm SPH}}$;\ \ \ \ \ 
$1<\rho_{\scriptstyle{\rm SPH}}/\langle\rho\rangle<10$ 
\item medium density regions, DTFE larger than SPH: \\ 
\hspace*{0.0cm} $\rho_{\scriptstyle{\rm DTFE}}\,>\,\rho_{\scriptstyle{\rm SPH}}$;\ \ \ \ \
$1<\rho_{\scriptstyle{\rm SPH}}/\langle\rho\rangle<10$
\item high density regions: \\
\hspace*{0.0cm} $\rho_{\scriptstyle{\rm DTFE}}\,\gtrsim\,\rho_{\scriptstyle{\rm SPH}}$;  
\ \ \ \ \ $\rho_{\scriptstyle{\rm SPH}}/\langle\rho\rangle\,>\,10$ 
\end{enumerate}
\noindent The physical meaning of the distinct sectors in the scatter diagram 
becomes apparent when relating the various regimes with the spatial
distribution of the corresponding particles. This may be appreciated
from the five subsequent frames in Fig.~2, each depicting the related
particle distribution in the same slice of width 0.04 kpc. The centre
and bottom frames, numbered 1 to 4, show the spatial distribution of
each group of particles, isolated from the complete distribution (top
right frame, Fig.~2). These particle slices immediately reveal the
close correspondence between any of the sectors in the scatter diagram
and typical features in the spatial matter distribution of the
two-phase interstellar medium. This systematic behaviour seems to
point to truly fundamental differences in the workings of the SPH and
DTFE methods, and would be hard to understand in terms of random
errors. The separate spatial features in the gas distribution seem to
react differently to the use of the DTFE method.

We argue that the major share of the disparity between the SPH and
DTFE density estimates has to be attributed to SPH, mainly on the
grounds of the known fact that SPH is poor in handling nontrivial
configurations such as encountered in multiphase media. By separately
assessing each regime, we may come to appreciate how these differences
arise. In sector 1, involving the diffuse low density medium, the DTFE
and SPH estimates are of comparable magnitude, be it that we do
observe a systematic tendency.  In the lowest density realms, whose
relatively smooth density does not raise serious obstacles for either
method, DTFE and SPH are indeed equal (with the exception of
variations to be attributed to random noise). However, near the edges
of the low density regions, SPH starts to overestimate the local
density as the kernels do include particles within the surrounding
high density structures. The geometric interpolation of the DTFE
manages to avoid this systematic effect (see e.g.~Schaap \& van de
Weygaert 2002a,b), which explains the systematic linear decrease of
the ratio $\rho_{\scriptstyle{\rm DTFE}}/\rho_{\scriptstyle{\rm SPH}}$
with increasing $\rho_{\scriptstyle{\rm SPH}}/\langle\rho\rangle$. To
the other extreme, the high density regions in sector 4 are identified
with compact dense clumps as well as with their extensions into
connecting filaments and walls. On average DTFE yields higher density
estimates than SPH, frequently displaying superior spatial resolution
(see also greyscale plot in Fig.~1). Note that the repercussions may
be far-reaching in the context of a wide variety of astrophysical
environments characterized by strongly density dependent physical
phenomena and processes ! The intermediate regime of sectors 2 and 3
clearly connects to the filamentary structures in the gas
distribution. Sector 2, in which the DTFE estimates are larger than
those of SPH, appears to select out the inner parts of the filaments
and walls. By contrast, the higher values for the SPH produced
densities in sector 3 are related to the outer realms of these
features. This characteristic distinction can be traced back to the
failure of the SPH procedure to cope with highly anisotropic particle
configurations. While it attempts to maintain a fixed number of
neighbours within a spherical kernel, it smears out the density in a
direction perpendicular to the filament. This produces lower estimates
in the central parts, which are compensated for with higher estimates
in the periphery. Evidently, the adaptive nature of DTFE does not
appear to produce similar deficiencies.

\section{The DTFE particle method}
Having demonstrated the improvement in quality of the DTFE density
estimates, this suggests a considerable potential for incorporating
the DTFE in a self-consistent manner within a hydrodynamical code.
Here, we first wish to indicate a possible route for accomplishing
this in a particle hydrodynamics code through replacement of the
kernel based density estimates (\ref{sphdensity}) by the DTFE density
estimates. We are currently in the process of implementing this.  The
formalism on which this implementation is based can be easily derived,
involving nontrivial yet minor modifications. Essentially, it uses the
same dynamic equations for gas particles as those in the regular SPH
formalism, the fundamental adjustment being the insertion of the DTFE
densities instead of the regular SPH ones. In addition, a further
difference may be introduced through a change in treatment of viscous
forces. Ultimately, this will work out into different equations of
motion for the gas particles. A fundamental property of a DTFE based
hydrocode, by construction, is that it conserves mass exactly 
and therefore obeys the continuity equation. This is not necessarily
true for SPH implementations (Hernquist \& Katz \cite{HK89}).

The start of the suggested DTFE particle method is formed by the
discretized expression for the Lagrangian $L$ for a compressible,
nondissipative flow,
\begin{equation}
\label{gaslagrangian} 
L\,=\,\sum_i m_i \left( \frac{1}{2} v_i^2 + u_i(\rho_i,s_i) \right)\,,
\end{equation}
\noindent where $m_i$ is the mass of particle $i$, $v_i$ its velocity, $s_i$ the 
corresponding entropy and $u_i$ its specific internal energy. In this
expression, $\rho_i$ is the density at location $i$, as yet
unspecified.  The resulting Euler-Lagrange equations are
\begin{equation}
\label{eqmotion1}
\frac{dv_i}{dt}\,=\,-\sum_j m_j \left(\frac{\partial u_j}{\partial {\rho}_j}\right)_{\!\!s}\, 
 \frac{\partial \rho_j}{\partial x_i}\,.
\end{equation}
\noindent The standard SPH equations of motion then follow after inserting the SPH density 
estimate (Eq.~\ref{sphdensity}). Instead, insertion of the DTFE
density (Eq.~\ref{DTFEdensity}) will lead to the corresponding
equations of motion for the DTFE-based formalism. Note that the usual
conservation properties related to Eq.~\ref{gaslagrangian} remain
intact. After some algebraic manipulation, thereby using the basic
thermodynamic relation for a gas with equation of state $P(\rho)$,
\begin{equation}
\left(\frac{\partial u_i}{\partial \rho_i} \right)_{\!\!s}\,=\,\frac{P_i}{\rho_i^2}\,,
\end{equation}
\noindent we finally obtain the equations of motion for the gas particles (moving 
in $D$-dimensional space),
\begin{equation}
\label{eqmotion2}
\frac{d v_i}{d t}\,=\,\frac{1}{D+1}\,\sum_{m=1}^{N_T} P({T_{m,i}})\,
\frac{\partial {\cal V}(T_{m,i})}{\partial x_i} \,.
\end{equation}
\noindent This expression involves a summation over all $N_T$ Delaunay tetrahedra
$T_{m,i}$, with volumes ${\cal V}(T_{m,i})$, which have the particle
$i$ as one of its four vertices. The pressure term $P({T_{m,i}})$ is
the sum over the pressures $P_j$ at the four vertices $j$ of
tetrahedron $T_{m,i}$, $P({T_{m,i}})=\sum P_j$.

As an interesting aside, we point out that unlike in the conventional
SPH formalism, this procedure implies an exactly vanishing
acceleration $dv_i/dt$ in the case of a constant pressure $P$ at each
of the vertices of the Delaunay tetrahedra containing particle $i$ as
one of their vertices. The reason for this is that one can then invoke
the definition of the volume of the contiguous Voronoi cell
corresponding to point $i$ (Eqn.~\ref{eq: contigvor}), yielding
\begin{equation}
 \frac{d v_i}{d t}\,=\,\frac{1}{D+1}\,P \,
\frac{\partial {\cal W}_{Vor,i}}{\partial x_i} \,.
\end{equation}
\noindent Since the volume of the contiguous Voronoi cell does not depend on the
position of particle $i$ itself (it lies in the interior of the
contiguous Voronoi cell), the resulting acceleration vanishes. Another
interesting notion, which was pointed out by Icke (\cite{ick2002}), is
that Delaunay tessellations also provide a unique opportunity to
include a natural treatment of the viscous stresses in the physical
system. We intend to elaborate on this possibility in subsequent work
dealing with the practical implementation along the lines sketched
above.

\section{Delaunay tessellations\\
  \hspace*{0.3cm} and `moving grid' hydrocodes} Ultimately, the ideal
hydrodynamical code would combine the advantages of the Eulerian as
well as of the Lagrangian approach. In their simplest formulation,
Eulerian algorithms cover the volume of study with a fixed grid and
compute the fluid transfer through the faces of the (fixed) grid cell
volumes to follow the evolution of the system. Lagrangian
formulations, on the other hand, compute the system by following the
ever changing volume and shape of a particular individual element of
gas (interestingly, the `Lagrangian' formulation is also due to Euler
\cite{eul1862}, who employed this formalism in a letter to Lagrange,
who later proposed these ideas in a publication by himself,
\cite{lag1762}; see Whitehurst~\cite{Wh95}). 

For a substantial part the success of the DTFE may be ascribed to the 
use of Delaunay tessellations as an optimally covering grid. This 
suggests that they may also be ideal for the use in moving grid 
implementations for hydrodynamical calculations. As in our SPH 
application, such hydrocodes with Delaunay tessellations at their 
core would warrant a close connection to the underlying matter 
distribution. Indeed, attempts towards such implementations have already 
been introduced in the context of a few specific, mainly two-dimensional, 
applications (Whitehurst~\cite{Wh95}, Braun \& Sambridge~\cite{BraSam95},
Sukumar~\cite{Suk98}). Alternative attempts towards the development of
moving grid codes, in an astrophysical context, have shown their
potential (Gnedin \cite{Gne95}, Pen \cite{Pen98}).

For a variety of astrophysical problems it is indeed essential to have
such advanced codes at one's disposal. An example of high current
interest may offer a good illustration. Such an example is the
reionization of the intergalactic medium by the ionizing radiation
emitted by the first generation of stars, (proto)galaxies and/or
active galactic nuclei. These radiation sources will form in the
densest regions of the universe. To be able to resolve these in
sufficient detail, it is crucial that the code is able to focus in
onto these densest spots.  Their emphasis on mass resolution makes
Lagrangian codes --~including SPH~-- usually better equipped to do so,
be it not yet optimally. On the other hand, it is in the low density
regions that most radiation is absorbed at first. In the early stages
the reionization process is therefore restricted to the huge underdense
fraction of space. Simulation codes should therefore properly represent
and resolve the gas density distribution within these voidlike regions. 
The uniform spatial resolution of the Eulerian codes is better suited to
accomplish this. Ideally, however, a simulation code should be able to
combine the virtues of both approaches, yielding optimal mass resolution
in the high density source regions and a proper coverage of the large
underdense regions. Moving grid methods, of which Delaunay tessellation 
based ones will be a natural example, may indeed be the best alternative,
 as the reionization simulations by Gnedin (\cite{Gne95}) appear to indicate.
 There have been many efforts in the context of Eulerian codes towards 
the development of Adaptive Mesh Refinement( AMR) algorithms(
Berger~\cite{BC89}), which have achieved a degree of maturity. Their chief
advantage is their ability to concentrate computational effort on regions 
based on arbitrary refinement criteria, where, in the basic form at least,
moving grid methods refine on a mass resolution criterion. However they are
still constrained by the use of regular grids, which may introduce artifacts
due to the presence of preferred directions in the grid.
The advantages of a moving grid fluid dynamics code based on Delaunay
tessellations have been most explicitly demonstrated by the
implementation of a two-dimensional lagrangian hydrocode (FLAME) by
Whitehurst (\cite{Wh95}). These advantages will in principle apply to
any such algorithm, in particular also for three-dimensional
implementations (of which we are currently unaware). Whitehurst
(\cite{Wh95}) enumerated various potential benefits in comparison with
conventional SPH codes, most importantly the following:
\begin{enumerate}
\item    SPH needs a smoothing length h. 
\item    SPH needs an arbitrary kernel function W. 
\item    The moving grid method does not need an (unphysical) artificial
  viscosity to stabilize solutions. 
\end{enumerate}
\noindent The validity of the first two claims has of course also been 
demonstrated in this study for particle methods based on DTFE.
Whitehurst showed additionally that there is an advantage of
moving grid methods over Eulerian grid-based ones. The implementation
of Whitehurst, which used a first-order solver 
and a limit on the shape of grid cells to control the effects of 
shearing of the grid, was far superior to
all tested first-order Eulerian codes, and superior to many
second-order ones as well. The adaptive nature of the Langrangian
method and the fact that the resulting grid has no preferred
directions are key factors in determining the performance of moving
grid methods such as FLAME. For additional convincing arguments,
including the other claims, we may refer the reader to the truly
impressive case studies presented by Whitehurst (\cite{Wh95}).

\section{Summary \& discussion}

Here we have introduced the DTFE as an alternative density estimator
for particle fluid dynamics. Its principle asset is that it is fully
self-adaptive, resulting in a density field reconstruction which
closely reproduces, usually in meticulous detail, the characteristics
of the spatial particle distribution. It may do so because of its
complete independence of arbitrary user-specified smoothing functions
and parameters. Unlike conventional
methods, such as the kernel estimators used in SPH, it manages to
faithfully reproduce the anisotropies in the local particle
distribution. It therefore automatically reflects the genuine geometry
and shape of the structures present in the underlying density field.
This is in marked contrast with kernel based methods, which almost
without exception produce distorted shapes of density features, the
result of the convolution of the real structure with the intrinsic
shape of the smoothing function. Its adaptive and local nature also
makes it optimally suited for reconstructing the hierarchy of scales
present in the density distribution. In kernel based methods the
internal structural richness of density features is usually suppressed
on scales below that of the characteristic (local) kernel scale. DTFE,
however, is solely based upon the particle distribution itself and 
follows the density field wherever the discrete
representation by the particle distribution allows it to do so. Its
capacity to resolve structures over a large dynamic range may prove to
be highly beneficial in many astrophysical circumstances, quite often 
involving environments in which we encounter a hierarchical embedding 
of small-scale structures within more extended ones.

In this study we have investigated the performance of the DTFE density
estimator in the context of a Smooth Particle Hydrodynamics simulation
of a multiphase interstellar medium of neutral gas.  The limited
spatial resolution of current particle hydrodynamics codes are known
to implicate considerable problems near regions with e.g.~steep density
and temperature gradients. In particular their handling of shocks
forms a source of considerable concern. SPH often fails in and
around these regions, so often playing a critical and vital role
in the evolution of a physical system. Our study consists of a
comparison and confrontation of the conventional SPH kernel based
density estimation procedure with the corresponding DTFE density field
reconstruction method.

The comparison of the density field reconstructions demonstrated
convincingly the considerable improvement embodied by the DTFE
procedure. This is in particular true at locations and under
conditions where SPH appears to fail. Filamentary and sheetlike
structures provide telling examples of the superior DTFE handling with
respect to the regular SPH method, with the most pronounced
improvement occurring in the direction of the steepest density
gradient.

Having shown the success of the DTFE, we are convinced that its application
towards the analysis of the outcome of SPH simulations will prove to
be highly beneficial. This may be underlined by considering a fitting
illustration.  Simulations of the settling and evolution of the X-ray
emitting hot intracluster gas in forming clusters of galaxies do
represent an important and cosmologically relevant example (see
Borgani \& Guzzo \cite{BorGuz01} and Rosati et al.~\cite{Rosetal02}
for recent reviews). The X-ray luminosity is strongly dependent upon
the density of the gas. The poor accuracy of the density determination
in regular SPH calculations therefore yields deficient X-ray
luminosity estimates (see Bertschinger \cite{Ber98} and Rosati et
al.~\cite{Rosetal02} for relevant recent reviews). Despite a number of
suggested remedies, such as separating particles according to their
temperature, their ad-hoc nature does not evoke a strong sense of
confidence in the results. Numerical limitations will of course always
imply a degree of artificial smoothing, but by invoking tools based
upon the DTFE technique there is at least a guarantee of an optimal
retrieval of information contained in the data.

Despite its promise for the use in a variety of analysis tools for
discrete data samples, such as particle distributions in computer simulations
or galaxy catalogues in an observational context, its potential would be most optimally
exploited by building it into genuine new fluid dynamics codes. Some
specific (two-dimensional) examples of succesful attempts in other
scientific fields were mentioned, and we argue for a similar strategy
in astrophysics.  One path may be the upgrade of current particle
hydrodynamics codes by inserting DTFE technology. In this study, we
have outlined the development of such a SPH-like hydrodynamics scheme
in which the regular kernel estimates are replaced by DTFE estimates.
One could interpret this in terms of the replacement of the
user-specified kernel by the self-adaptive contiguous Delaunay cell,
solely dependent on the local particle configuration.  An
additional benefit will be that on the basis of the localized
connections in a Delaunay tessellations it will be possible to define
a more physically motivated artificial viscosity term.

The ultimate hydrodynamics algorithm would combine the virtues of
Eulerian and Lagrangian techniques. Considering the positive
experiences with DTFE, it appears to be worthwhile within the context
of `moving grid' or `Lagrangian grid' methods to investigate the use
of Delaunay tessellations for solving the Euler equations. With
respect to a particle hydrodynamics code, the self-adaptive virtues of
DTFE and its ability to handle arbitrary density jumps with only one
intermediate point may lead to significant improvements in the
resolution and shock handling properties. Yet, for grid based 
methods major complications may be expected in dealing with the
non-regular nature of the corresponding cells, complicating the
handling of flux transport along the boundaries of the
Delaunay tetrahedra.

The computational cost of DTFE resembling techniques is not
overriding. The CPU time necessary for generating the Delaunay
tessellation corresponding to a point set of $N$ particles is in the
order of $O(N\,{\rm log}\,N)$, comparable to the cost of generating
the neighbour list in SPH. Within an evolving point distribution
these tessellation construction procedures may be made far more
efficient, as small steps in the development in the system will induce
a correspondingly small number of tetrahedron (identity) changes. Such
dynamic upgrading routines are presently under development.

In summary, in this work we have argued for and demonstrated the potential and
promise of a natural computational technique which is based upon one
of the most fundamental and natural tilings of space, the Delaunay
tessellation. Although the practical implementation will undoubtedly
encounter a variety of complications, dependent upon the physical
setting and scope of the code, the final benefit of a natural moving
grid hydrodynamics code for a large number of astrophysical issues may
not only represent a large progress in a computational sense. Its
major significance may be found in its ability to address fundamental
astrophysical problems in a new and truely natural way, leading to
important new insights in the workings of the cosmos.

\begin{acknowledgements}
  The authors thank Vincent Icke for providing the incentive for this
  project and for fruitful discussions and suggestions. WS is grateful
  to Emilio Romano-D\'{\i}az for stimulating discussions. RvdW thanks
  Dick Bond and Bernard Jones for consistent inspiration and encouragement
  to investigate astrohydro along tessellated paths.
  We are also indebted to Jeroen Gerritsen and Roelof Bottema for providing
   software and analysis tools. 
\\
 This work was sponsored by the stichting Nationale Computerfaciliteiten
(National Computing Facilities Foundation) for the use of supercomputer 
facilities, with financial support from the Nederlandse Organisatie voor
Wetenschappelijk Onderzoek (Netherlands Organisation for Scientific Research,
NWO)

\end{acknowledgements}

\end{document}